\documentclass[a4paper,fleqn,usenatbib]{mnras}

\usepackage[T1]{fontenc}
\usepackage{ae,aecompl}

\usepackage{graphicx}
\usepackage{amsmath}
\usepackage{amssymb}
\usepackage{tabularx}

\title[]{Linking Long- and Short-Term Emission Variability in Pulsars}

\author[P. R. Brook et al.]{
  P. R. Brook,$^{1,2}$\thanks{E-mail: paul.brook@gmail.com}
  A. Karastergiou,$^{3,4,5}$
  and S. Johnston,$^{6}$
  \\
  $^{1}$Department of Physics and Astronomy, West Virginia University,
  Morgantown, WV 26506, USA\\
  $^{2}$Center for Gravitational Waves and Cosmology, West Virginia
  University, Chestnut Ridge Research Building, Morgantown, WV 26505,
  USA\\
  $^{3}$Astrophysics, University of Oxford, Denys Wilkinson
  Building, Keble Road, Oxford, OX1 3RH, UK\\
  $^{4}$Physics Department, University of the Western Cape, Cape Town
  7535, South Africa\\
  $^{5}$Department of Physics and Electronics, Rhodes University, PO
  Box 94, Grahamstown 6140, South Africa\\
  $^{6}$CSIRO Astronomy and Space Science, Australia Telescope
  National Facility, P.O. Box 76, Epping, NSW 1710, Australia\\
}

\date{Accepted XXX. Received YYY; in original form ZZZ}

\pubyear{2019}

\begin{document}
\label{firstpage}
\pagerange{\pageref{firstpage}--\pageref{lastpage}}
\maketitle

\begin{abstract}
  It is now known that the emission from radio pulsars can vary over a
  wide range of timescales, from fractions of seconds to
  decades. However, it is not yet
  known if long- and short-term emission variability are caused by the
  same physical processes. It has been observed that
  long-term emission variability is often correlated with rotational
  changes in the pulsar. We do not yet know if the same is true of
  short-term emission variability, as the rotational changes involved
  cannot be directly measured over such short timescales.
  To remedy this, we propose a continuous pulsar monitoring technique
  that permits the statistical detection
  of any rotational changes in nulling and mode-changing pulsars with
  certain properties. Using a simulation, we explore the range of
  pulsar
  properties over which such an experiment would be possible.

\end{abstract}

\begin{keywords}
  pulsars: general -- pulsars: individual: J1701-3726 -- pulsars:
  individual: J1727-2739 -- stars: neutron
\end{keywords}

\section{Introduction}
\label{intro}
Although individual radio pulses received from a pulsar can vary
substantially in phase and amplitude, the average of thousands of
pulses (the \emph{pulse profile})
is often considered stable and unique to each pulsar at a given
observational frequency. However, in contrast to this high degree of
emission
stability, some pulsars are observed to show variability on timescales
ranging from the order of a pulse period to many years.
In the early 1970s, it was discovered that emission changes can occur
in pulsars on short timescales, in the forms of \emph{nulling} and
\emph{mode-changing}
\citep{1970Natur.228...42B,1970Natur.228.1297B}. Mode-changing is a
phenomenon in which pulsars are seen to discretely switch between two
or more emission states. Nulling can be thought of as an extreme form
of mode-changing, with one state showing no, or low emission. The
timescale of mode-changing and nulling ranges from a few pulse periods
to many hours or even days \citep{2007MNRAS.377.1383W}. The fraction
of time in which the pulsar is in a null state (the \emph{nulling
  fraction}),
also varies from 0 to $\sim$ 95\%, and has been found to correlate
with both characteristic age \citep{1976MNRAS.176..249R} and pulse
period
\citep{1992ApJ...394..574B}.
\\
Rotating radio transients (\emph{RRATs}) are a class of pulsar which
produces
detectable emission (bursts typically lasting milliseconds) only
sporadically, at irregular and infrequent intervals, with nulls of
minutes to hours. The nulling fraction for RRATs can extend upwards of
99\%. More than 70 are now known since their discovery in 2006
\citep{2006Natur.439..817M}. Analysis of their burst arrival times
reveals underlying
regularity of the order of seconds and they have comparable spindown
rates to other neutron star classes.
\\
A group known as \emph{intermittent pulsars} go through a
quasi-periodic cycle between phases in which radio emission is and is
not detected \citep{2006Sci...312..549K,
  2012ApJ...746...63C,2012ApJ...758..141L,2017ApJ...834...72L}
with timescales of variability ranging from weeks
to years. The intermittent pulsar discovered by Kramer et al. was the
first example of a pulsar showing emission changes that were strongly
linked to rotational behaviour. In these objects, each of their two
states is associated with a distinct rate of rotational energy loss
(spindown rate $\dot{\nu}$).
Pulsars with nulls of many hours are sometimes seen as a bridge
between the nulling pulsars described above and intermittent
pulsars. Objects of this type are, therefore,
also sometimes labelled as intermittent pulsars
\citep[e.g.][]{2016MNRAS.456.3948H}. Throughout this paper, however,
we will use the term to describe only those pulsars which have
timescales of weeks or more and have confirmed emission-rotation
correlation.
\\
\citet{2010Sci...329..408L} showed six pulsars for which the spindown
rate
is also correlated with changes in emission (this time in the shape of
the pulse profile) over long timescales. PSR~J0742$-$2822 shows the
most rapid changes, switching on a timescale of around 100 days, while
PSR B2035+36 showed only 1 switch in 19 years of observation. Further
examples of this kind of \emph{state-switching} are seen in
\citet{2014ApJ...780L..31B} and \citet{2016MNRAS.456.1374B}.
An explanation for the emission-rotation correlation seen in some
pulsars was first proposed by \citet{2006Sci...312..549K}. They
suggested that changing currents of charged particles in the pulsar
magnetosphere are responsible for both emission changes and
variations in braking torque.
\\
There may be a continuum of variability in the pulsar
population. In an attempt to unify these assorted types of
variability,
we question whether common processes are responsible for the very
different timescales that we observe. As emission-rotation correlation
is seen in intermittent and state-switching pulsars, a step towards
answering this question would be taken by discovering whether
shorter-term emission changes are also correlated with rotational
changes; if nulling and mode-changing are also caused by changing
magnetospheric currents, then the braking torque acting on the pulsar
and,
consequently, its rotational behaviour must also be affected. Because
these phenomena occur on timescales much shorter than the duration
over which the spindown rate can be measured (typically weeks due to
the small rate of spindown in comparison to measurement
uncertainties), we currently remain somewhat agnostic regarding any connection. However, some connections between mode-changing, state-switching and $\dot{\nu}$ have been suggested by \citet{2010Sci...329..408L}. One focus of their paper is
PSR~B1822-09, a well known mode-changing pulsar. The amount
of time spent in each emission mode during an observation inevitably has
consequences for the shape of the resulting integrated pulse
profile. Lyne et al. show hints that this mode-changing fraction may
change gradually over time and that the resulting integrated pulse
profile shape changes may have some level of correlation with
$\dot{\nu}$. However, the relationship shown is not conclusive; the
Jodrell Bank observations featured in Lyne et al. are of short
duration (between 6 and 18 minutes) and are unevenly spaced. It is
not possible to obtain a comprehensive understanding of the
mode-changing behaviour with such observations. Additionally, the
clear and simple metrics used as proxies for pulse profile shape by
Lyne et al. are too elementary to describe how the profile shape is
changing in any detail. At least partially as a result of these issues,
the relationship between profile shape and $\dot{\nu}$ in
PSR~B1822-09 is ambiguous; we see that profile shape and $\dot{\nu}$
appear correlated at two epochs, but also that shape changes at
other times coincide with an apparently stable $\dot{\nu}$.
\\
In this paper, we propose a method that can potentially obtain
rotational information from a pulsar on timescales of a pulse period;
at present, no other technique permits the investigation of rotational
behaviour on such short timescales. The method can be used to
statistically infer whether mode-changing and nulling are accompanied
by a change in spindown rate and potentially allows us to take a step
towards unification in the domain of pulsar variability. In related
work, \citet{2018MNRAS.475.5443S} inject $\dot{\nu}$
transitions into simulated pulsar timing data and assess how reliably
they can recover the transition parameters. The ability to do so
depends on the pulse time of arrival (TOA) precision, the observing cadence, the number of
$\dot{\nu}$ transitions injected, their amplitude and the separation
time between them. We discuss their work in the context of this paper
in Section \ref{disc}.
\\
In Section \ref{cont} we describe how a continuous monitoring campaign
of a nulling or mode-changing pulsar could illuminate its rotational
behaviour on short timescales. We also outline a simulation of this
scenario in order to explore the range of parameters over which such
an experiment would be possible. In order to constrain these
parameters to within realistic boundaries, we have carried out
multiple observations of two nulling pulsars. The details of these
observations and their results are found in Section \ref{cons}.
In Section \ref{simu} we describe the results of the simulation based
on the nulling observations, and the findings are discussed in Section
\ref{disc}. Conclusions are drawn in Section \ref{conc}.
\section{Continuous Monitoring Proposal}
\label{cont}
The following is a scenario in which it would be possible to probe a
pulsar's
rotational behaviour on short timescales.
Consider a simple model of a pulsar that has two distinct emission
states, each
having a different rate of spindown. The pulsar switches between
states on timescales of minutes and hours. If we observe the
pulsar continuously for a span of time, we will know what fraction it
spent
in state A and what fraction in state B. These
\emph{state fractions} will be different for each observation span;
the degree to which they differ will depend on the length of the span
and the nature of the pulsar. If the observations are long enough, so
that an average spindown rate can be precisely measured, then it can
be demonstrated that two separate monitoring spans in which the
state fractions are different, would have a different average spindown
rate. In this way, the relationship between short-term emission
changes and pulsar rotation could be elucidated by continuous
monitoring of a sufficiently bright mode-changing or nulling pulsar;
an analysis of its emission will reveal the fraction of time spent in
each state over a certain duration. If we begin to see a correlation
between the fraction of time spent in an emission state and the
measured spindown rate, then we can infer that each emission state
also has a distinct spindown rate associated with it.
\\
We have created a simulation which models the behaviour
of a mode-changing or nulling (hereafter state-changing) pulsar,
produces artificial pulse TOAs and
we thereby test the range of parameters over which such as continuous
monitoring proposal could
be successful.
\subsection{Pulsar Simulation}
\label{simulation}
\begin{figure*}
  \centering
  \includegraphics[width=170mm]{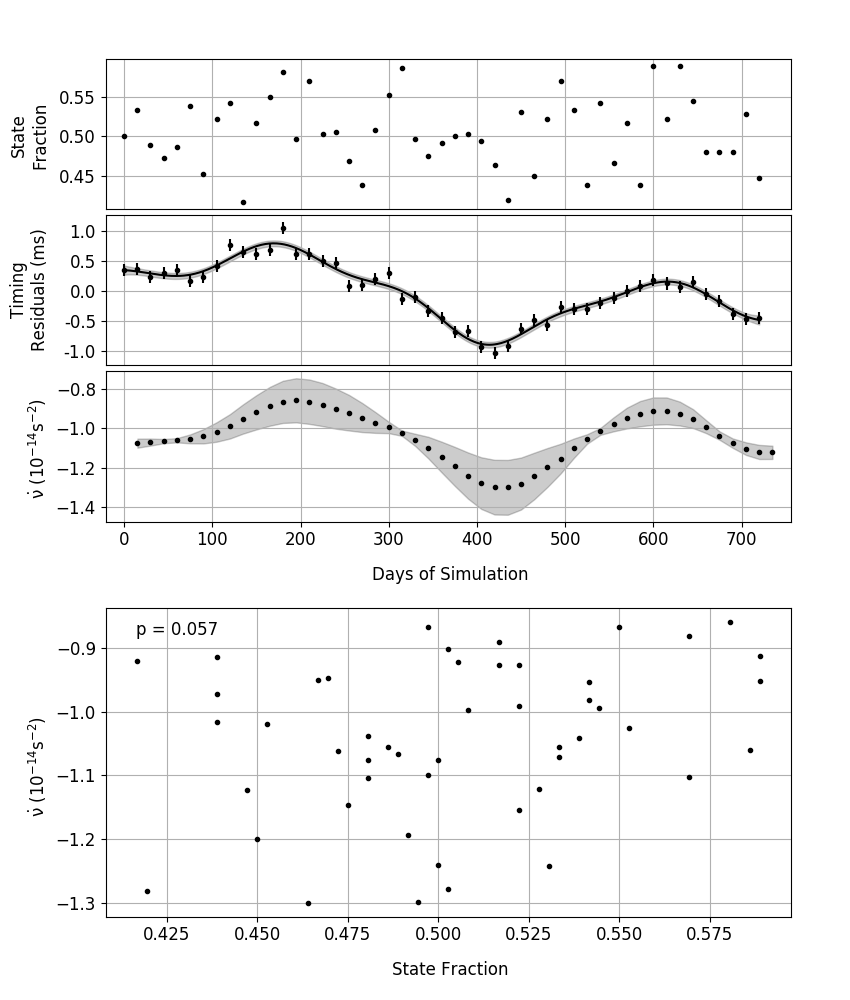}
  \caption[]{One particular realisation of the simulation in which the
    state fraction varies stochastically (see
    Section~\ref{stochastic}), with $\sigma_{\rm SF} = 0.03$ and
    $|\Delta \dot{\nu}| = 1 \times 10^{-14}$~s$^{-2}$. Measurements
    are made for two simulated years in 15-day increments. Top panel:
    How the state fraction of the simulated pulsar changes with
    time. Second panel down: The black dots are the timing residuals
    with respect to a timing model fit by TEMPO2
    \citep{2006MNRAS.369..655H} using the simulated TOAs. The TOA
    measurement uncertainties are 100~$\mu$s. The black line is a GP
    regression model fit to the timing residuals and the grey shading
    indicates the $1\sigma$ uncertainty. Third panel down: GP
    regression allows us to analytically model the second derivative
    directly from the timing residuals, giving $\dot{\nu}$ with
    associated fully Bayesian error estimation (see
    \citealt{2016MNRAS.456.1374B} for details). The $\dot{\nu}$ values
    are calculated only at points in time where a TOA is
    produced. Bottom panel: The correlation between the state fraction
    and $\dot{\nu}$ calculated throughout the simulation. The degree
    of correlation corresponds to a $p$-value of 0.057 in this
    realisation, meaning that there is a 5.7\% probability that
    intrinsically non-correlated data would show at least this level
    of correlation.}
  \label{process}
\end{figure*}

Expressed as a Taylor expansion, a pulsar's rotation frequency is
given by
\begin{equation}
  \nu(t) = \nu_{0} + \dot{\nu}_{0}(t-t_{0}) +
  \frac{1}{2}\ddot{\nu}_{0}(t-t_{0})^{2} + \ldots,
  \label{taylor}
\end{equation}
where the subscript 0 denotes the value of a variable at some
reference
epoch $t_{0}$.
Rotation frequency is also $\dot{N}$, where $N$ is the pulse
number. We can
integrate equation \ref{taylor} to show that
\begin{equation}
  N = N_{0} + \nu_{0}(t-t_{0}) + \frac{1}{2}\dot{\nu}_{0}(t-t_{0})^{2} +
  \frac{1}{6}\ddot{\nu}_{0}(t-t_{0})^{3} + \ldots,
  \label{eN}
\end{equation}
where $N_{0}$ is the pulse number at $t_{0}$. For the analysis in this paper, $N$ can be accurately
  approximated by the first three terms on the right-hand side of
  Equation 2, given that we can neglect the contribution from
  $\ddot{\nu}$ over the timescales involved in the simulation.
\\
As Equation \ref{eN} assumes a constant $\dot{\nu}_{0}$, when this value makes a step change (as the pulsar switches state), the equation must be re-initiated with updated values of $N_{0}$, $\nu_{0}$ and $\dot{\nu}_{0}$ in order to keep track of $N$.\\
We choose the effective TOA for the
simulated pulsar to be taken when $N$ has an integer value,
i.e. when the pulsar beam is pointing towards Earth.
The simulation calculates $N$ whenever we require a TOA to be generated.
The time for the pulsar to rotate so that $N$ reaches the next
integer value is easily computed and so this is added to the
time at which $N$ was calculated to produce a simulated TOA.
\\
The following simulation parameters can be adjusted:
\begin{itemize}
\item initial rotation frequency $\nu_{0}$ of the pulsar.
\item state fraction of the pulsar.
\item $\dot{\nu}$ value of each pulsar state.
\item total observation span.
\item duration over which a measurement of $\dot{\nu}$ and state
  fraction is made.
\item uncertainty of measurement of a pulse TOA.
\item timescale of pulsar state changes.
\end{itemize}
To add noise to the simulated TOAs, a sample is drawn from a Gaussian
distribution with a zero mean and a standard deviation
$\sigma_{\rm{TOA}}$ of $1 \times 10^{-4}$ seconds. This accounts for
the template-fitting errors primarily due to radiometer noise. The
$\sigma_{\rm{TOA}}$ level chosen is that expected from a pulsar with a
10~ms
pulse width $W$ observed with a signal-to-noise ratio $S/N$ of 100.
\begin{equation}
  \sigma_{\rm {TOA}} \simeq \frac{W}{S/N}.
  \label{toa_uncert}
\end{equation}
\\
At an interval determined by the user, the simulation reaches a crossroad, at which point the pulsar can remain in its current state
or switch to the other. This is determined by the generation of a
random number and weighted by the value of the underlying state
fraction at that point in the simulation. For example, if the state
fraction is 0.9, a random number generated between 0 and 1 will
dictate that the simulated pulsar continues in State A when it is less
than 0.9 and State B otherwise.
\\
The simulation produces artificial barycentric pulse arrival
times as often as desired. Gaussian process (GP) regression is then
used to
model the noisy simulated data and to consequently track the pulsar's
$\dot{\nu}$ value. This is done by combining the technique of
\citet{2016MNRAS.456.1374B} with the use of the GP regression
software \textit{george} \citep{2015ITPAM..38..252A}.
In order to optimise the $\dot{\nu}$ models calculated by george, each
accepted
model for a data set is actually comprised of the median values of 100
others. The uncertainty of an accepted GP model is
determined by taking the standard deviation at each point across the
100 contributing models. Unphysical outliers (those models in which
$\dot{\nu}$ is
positive at any point) are removed before the median and standard
deviation are calculated. The $\dot{\nu}$ values, produced by the
simulation and calculated by GP regression, can then be compared to
the
state fraction over several observation spans and see if there is any
correlation. Figure~\ref{process} shows an example of the process.
\\
We can say a priori, that we would expect no significant correlation
between $\dot{\nu}$ and state fraction if any change in $\dot{\nu}$ is
so small as to be immeasurable. This would arise in a pulsar with: (i)
states with rates of spindown that are too similar, (ii)
a state fraction that does not vary over a wide enough range or (iii)
a low $S/N$ pulse profile, leading to a large $\sigma_{\rm {TOA}}$.
With regards to (i), we already have some information regarding the
$\dot{\nu}$
difference ($|\Delta\dot{\nu}|$) in two-state pulsars. In the six
\citet{2010Sci...329..408L} pulsars, the change in
spindown between the two states is approximately between 1 and
10\%. In the
intermittent pulsars, a $|\Delta\dot{\nu}|$ value as high as 150\% has
been
recorded \citep{2012ApJ...746...63C}.
With regards to (ii), the long term behaviour of the state fraction of
mode-changing and nulling pulsars in this context is currently
unknown; published mode-changing and nulling fractions,
(i.e. the fraction of the observation duration in which a pulsar is in
an alternative emission mode) have typically been obtained
through single, long-duration observations (e.g. two hours for
\citealt{2007MNRAS.377.1383W}).
In order to learn more about the behaviour of the state fraction and,
therefore, realistically constrain and model it within the simulation,
we have conducted observations of two nulling pulsars.
\section{Constraining State Fraction Variation}
\label{cons}
\citet{2007MNRAS.377.1383W} present two-hour observations of 23
pulsars which show evidence of nulling and/or mode-changing
behaviour. All observations were made in March or June of 2004, and
for each nulling pulsar, a nulling fraction was calculated.
In order to learn how a pulsar's nulling fraction behaves on long
timescales, we observed two pulsars in 2014 that were featured in the
work by
Wang et al. We calculated their nulling fractions and compared them to
the
2004 observations to see if and how these
values had changed over the intervening decade.
The pulsars observed were PSRs~J1701$-$3726 and J1727$-$2739. As
can be seen in Wang et al., both of these pulsars
switch frequently between states of emission and nulling over their
two-hour observations. PSR~J1701$-$3726 also shows some mode-changing
behaviour.
Any information regarding the behaviour of nulling fraction on long
timescales can help us constrain parameters in the state-changing
simulation.
\subsection{Observations and Analysis}
Both the 2004 and the 2014 data were recorded with one of the Parkes
Digital Filterbank systems
(PDFB1/2/3/4) with a total bandwidth of 256 MHz in 1024
frequency channels. Radio frequency interference was removed
using median-filtering in the frequency domain
then manually excising bad sub-integrations. Flux densities
have been calibrated by comparison to the continuum radio
source 3C 218. The data were then polarisation calibrated
for both differential gain and phase, and for cross coupling
of the receiver. The MEM method based on long observations
of PSR~J0437-4715 was used to correct for cross coupling
\citep{2004ApJS..152..129V}. Flux calibrations from Hydra A were
used to further correct the bandpass. After this calibration,
profiles were formed of total intensity (Stokes I),
and averaged over frequency. PSRs~J1701$-$3726 and J1727$-$2739 were
observed by Wang et
al. for two hours each on 20 March 2004. We observed PSRs~J1701$-$3726
for two hours on 2014 March 31 and two hours on 2014 April
02. We observed PSR~J1727$-$2739 for two hours on 2014 April 01
and two hours on 2014 April 03. In each observation around 2900 and
5500 single pulses were recorded from PSRs~J1701$-$3726 and
J1727$-$2739 respectively. GP regression was used
to model and subsequently flatten the baseline for each single pulse.
\subsection{Nulling Fraction Calculation Method}
\label{nf_calc}
In order to calculate nulling fraction, \citet{2007MNRAS.377.1383W} compare flux density histograms for windows centered (i) at pulse phases where the pulsar radio emission occurs and (ii) at phases far from the emission, where only noise is present. When comparing histograms, Wang et al. only consider bins that contain negative flux density values. One group of histogram bins are scaled until comparable with the other; the scaling factor provides the nulling fraction. Often, the best fit after scaling is still poor (depending on S/N and number of null pulses in the histogram). We have opted for a different technique which makes use of all of the flux density information centered at the phase of emission (further details in the following). In any case, more than the absolute nulling fraction, we are interested in seeing how much the fraction changes over time. This should be reflected similarly by both techniques.
\\
For each of the two observed pulsars, we measure the nulling
fraction in the following way. By looking at the integrated pulse
profile, we can define a phase window which contains all radio
emission
from the pulsar (see the top panels of Figures~\ref{1701_hist} and
\ref{1727_hist}). For each single pulse in an observation, we sum the
flux density in this phase window. A histogram of the summed flux
density values is then constructed for each pulsar observation (see
the bottom panels of Figures~\ref{1701_hist} and
\ref{1727_hist}). Each histogram analysed in this work appears to be
either bimodal or trimodal, showing one population of nulls and one or
two others of emission in the phase window. There is some
overlap between these populations; in order to disentangle them, we
fit the sum of multiple components to each histogram using non-linear
least squares. The null pulses are modelled by a Gaussian
component, and the emitting pulses by one or two log-normal components
\citep{2012MNRAS.423.1351B}.
We can calculate the nulling fraction
\begin{equation}
  NF = \frac{A_{n}}{A_{n} + A_{e}},
  \label{nf_eq}
\end{equation}
where $A_{n}$ and $A_{e}$ are the areas under the nulling and emission
distributions respectively.
The area of each component is given by $\sqrt{2\pi}\sigma_{d} H$,
where $\sigma_{d}$ is the standard deviation of each
distribution and $H$ is its height.
The function fitting uncertainty in the nulling fraction is found by
propagating the
uncertainties of the Gaussian and log-normal component parameters
($\sigma_{d}$ and $H$), as found by the
non-linear least squares fits to the histogram.
\subsection{PSR~J1701$-$3726}
\begin{figure*}
  \centering
  \includegraphics[width=190mm]{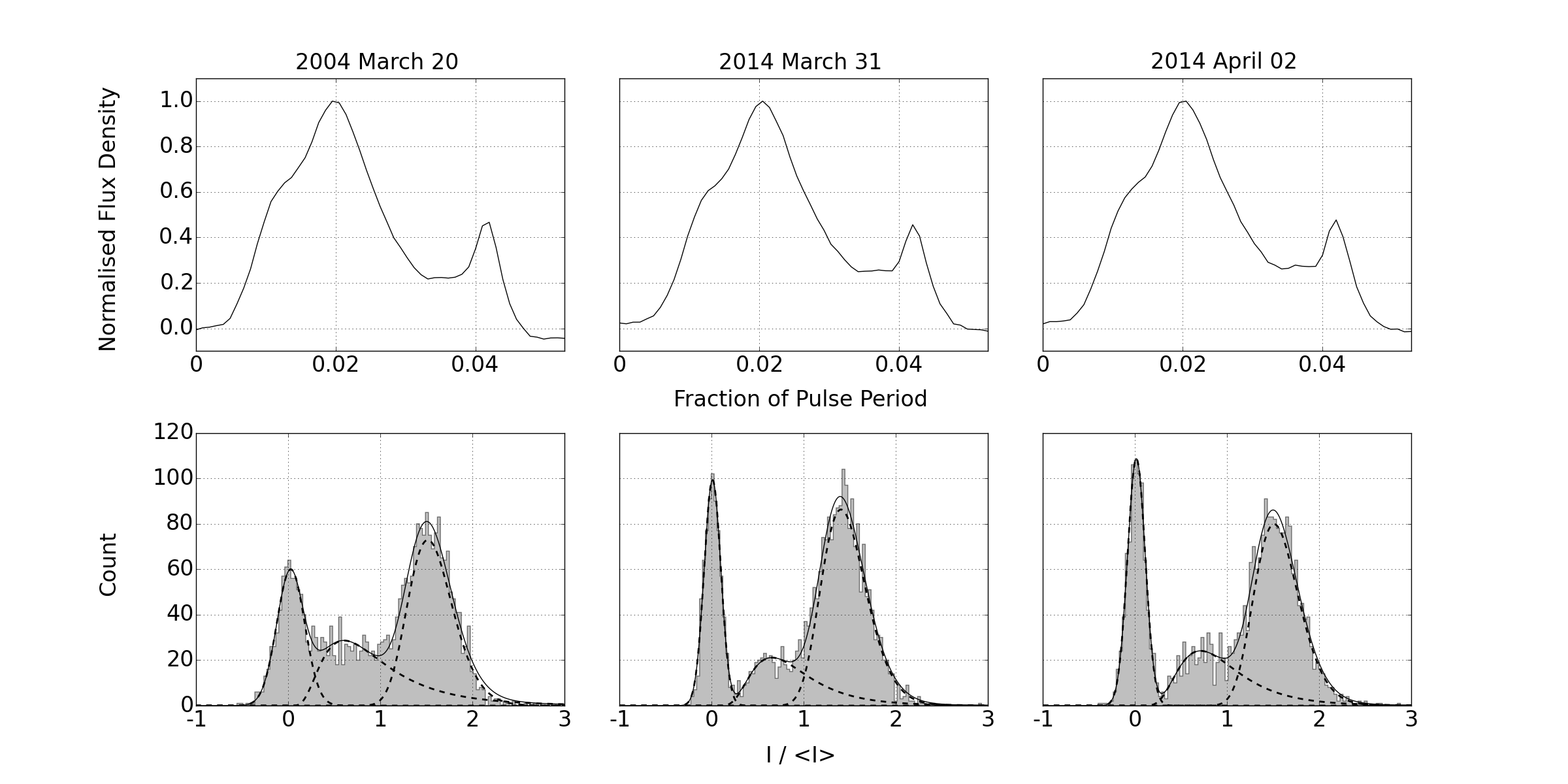}
  \caption[]{The pulse profiles and emission histograms for three
    observations of PSR~J1701$-$3726. The dates of the
    observations
    are shown at the top. Top panels: The flux density
    integrated over the whole two-hour observation. The
    section of the
    pulse profile shown is the phase window chosen to
    encompass all
    emission from the pulsar (see Section
    \ref{nf_calc} for details). Bottom panels:
    Histogram showing the total flux density
    summed across the
    emission window (intensity $I$, normalised
    by mean intensity <$I$>) for each
    single pulse ($\sim$ 2900 in
    total). The dashed lines are Gaussian
    and log-normal functions that are
    fit to the population of nulls and
    two
    populations of pulses
    respectively. The solid line shows the sum of the dashed lines.}
  \label{1701_hist}
\end{figure*}

PSR~J1701$-$3726 is both a nulling and mode-changing pulsar. During a
two-hour observation, \citet{2007MNRAS.377.1383W} observed the pulsar
to spend the
majority of its time in a mode where a trailing edge profile is much
smaller than the rest of the pulse, and a rarer mode in which the
pulse profile displays two peaks of
roughly equal height. These two emission modes are punctuated by
frequent and short nulling periods; the emission variability
occurs on minute timescales.
The emission windows and the flux density integrated over the whole
observation is depicted in the top row of Figure~\ref{1701_hist} for
three observations (one from 2004 and two from 2014). Each panel in
the bottom row of the figure shows a histogram for values of the flux
density summed over the emission window for each single pulse.
It is possible that the nulling state of PSR~J1701$-$3726, and each of
the two emission states may all have distinct spindown rates and so
would
not be a good candidate for a continuous monitoring campaign described
in this work. However, the observations separated by a decade may
still provide useful information regarding the behaviour of nulling
fraction on long timescales.
Using the distributions fitted to each histogram population
and Equation \ref{nf_eq}, we present the calculated nulling fractions
for
the three observation of PSR~J1701$-$3726 in Table~\ref{nf_table}.
\subsection{PSR~J1727$-$2739}
\begin{figure*}
  \centering
  \includegraphics[width=190mm]{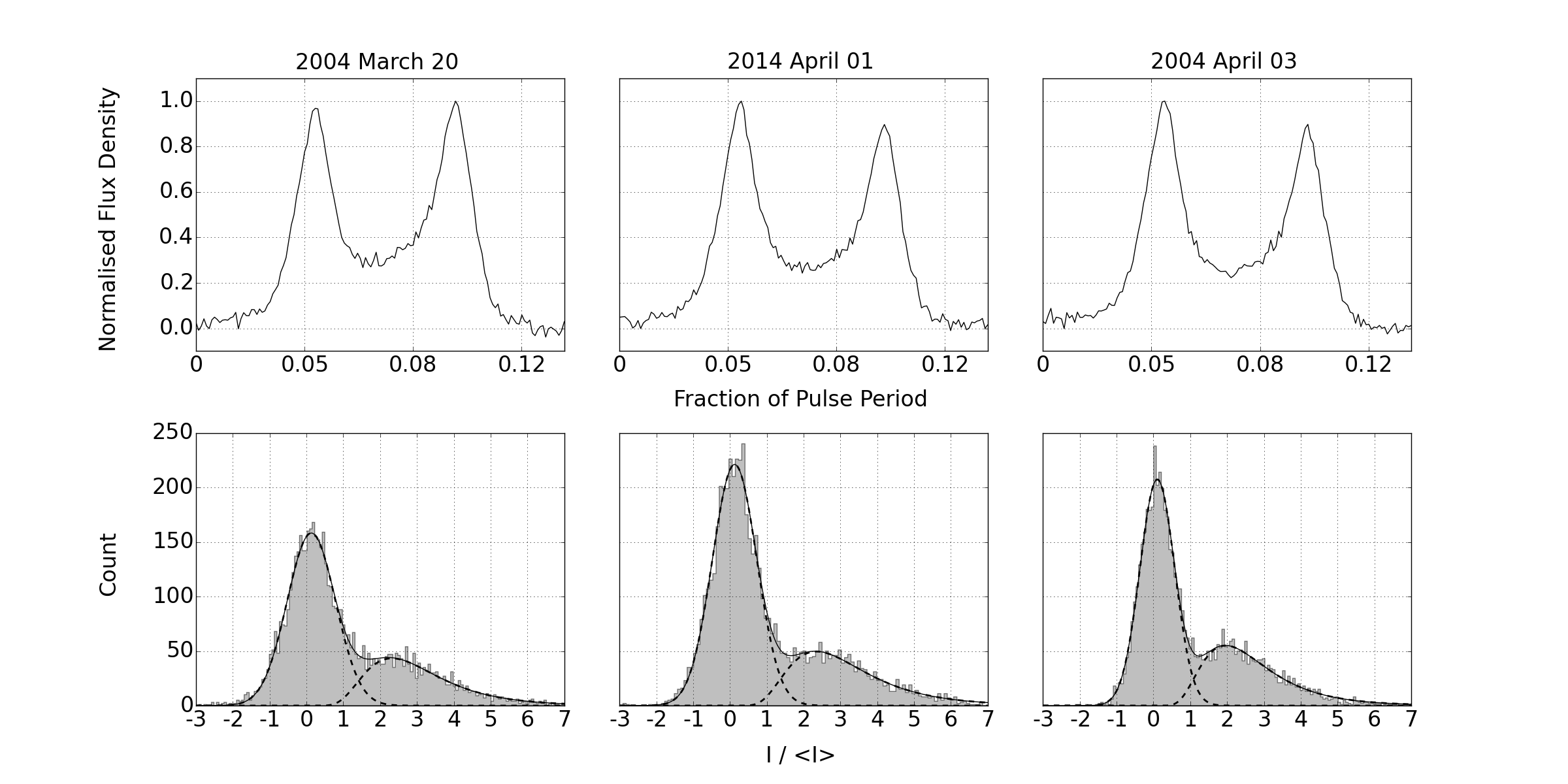}
  \caption[]{The pulse profiles and emission histograms for three
    observations of PSR~J1727$-$2739. The two-hour observation
    consisted of $\sim$ 5500 single pulses. The dashed lines are
    Gaussian and log-normal functions that are fit to a population
    of nulls and a
    population of pulses respectively. As
    Figure~\ref{1701_hist} otherwise.}
  \label{1727_hist}
\end{figure*}

PSR~J1727$-$2739 is a nulling pulsar that shows no signs of
mode-changing. \citet{2007MNRAS.377.1383W} report that the pulsar
emits frequent short bursts separated
by null intervals, and that this emission variability occurs on minute
timescales.
The observation-integrated pulse profiles as seen in the top row of
Figure~\ref{1727_hist} show a double peak, with each
component being of comparable height. The relative height is not
constant in all of observations; for the 2004 observation, the flux
density level is highest in the trailing peak, in contrast to the 2014
observations. The bottom row of Figure~\ref{1727_hist} shows a bimodal
histogram for
each observation, depicting a population of nulls and one of pulses.
The nulling fraction for PSR~J1727$-$2739, calculated using
Equation~\ref{nf_eq} is
shown in Table~\ref{nf_table}.

\subsection{Nulling Fraction Results}

\begin{table}
  \centering
  \caption{The calculated nulling fractions for three observations,
    each of two-hour duration, for PSRs~J1701$-$3726 and
    J1727$-$2739.}
  \label{nf_table}
  \begin{tabular}{cc} 
    \multicolumn{2}{c}{PSR~J1701$-$3726}\\
    \hline
    Observation Date & Nulling Fraction\\
    \hline
    2004/03/20 & 23.4 $\pm$ 5.0\%\\ 
    2014/03/31 & 24.2 $\pm$ 4.3\%\\ 
    2014/04/02 & 27.2 $\pm$ 4.6\%\\ 
    \hline\\[0.5cm]
    \multicolumn{2}{c}{PSR~J1727$-$2739}\\
    \hline
    Observation Date & Nulling Fraction\\
    \hline
    2004/03/20 & 51.7 $\pm$ 7.0\%\\ 
    2014/04/01 & 57.4 $\pm$ 7.1\%\\ 
    2014/04/03 & 55.6 $\pm$ 6.8\%\\ 
    \hline
  \end{tabular}
\end{table}
The limited data we have do not reveal any significant changes in the
nulling fraction of either observed pulsar after measurement
uncertainties are taken into consideration. Additionally, even if the
underlying nulling fraction does not change, we still expect
statistical fluctuations during a two-hour observation. To approximate
these, we can model the pulsar emission as a binomial process. Each of
our pulsars switches between states of null and emission on roughly
minute timescales; a two-hour observation would mean that the number
of trials $n$ = 120. We can take the average of our three observations
to find the probability $p$ of the pulsar being in a nulling
state. For PSR~J1701$-$3726, $p$ = 0.23; the standard deviation of the
number of nulls in a two-hour observation
$\sigma_{n}=\sqrt{np\left(1-p\right)} = 4.6$. The standard deviation
of the nulling fraction for a two-hour PSR~J1701$-$3726 observation,
therefore, is $\sim$ 4.6/120 = 3.8\%. For PSR~J1727$-$2739, $p$ = 0.60
and $\sigma_{n} = 5.4$. The standard deviation of the nulling fraction
for a two-hour PSR~J1727$-$2739 observation is 4.5\%. In
Table~\ref{nf_table} these statistical uncertainties are added in
quadrature to the distribution fitting uncertainties. For a 15-day
observation (around the length required for a precise measurement of
$\dot{\nu}$), the statistical uncertainty of the nulling fraction
drops to just $\sim$ 0.3\% for both pulsars. We cannot be
sure that the changes in nulling fraction that we observe are entirely
statistical or due to fitting uncertainties, and not caused by
a change (at least in part) in a  physical process intrinsic to the
pulsar. We consider all of this information when running the
state-changing simulation described in the next section.
\section{Simulation Results}
\label{simu}
We simulate a state-changing pulsar with a variety of parameters in
order to explore the parameter space over which it would be feasible
to detect distinct values of $\dot{\nu}$ in each state. The parameters
and their simulated values are summarised in Table~\ref{sim_table}.
We focus on two different scenarios: one in which the
state fraction varies stochastically around a certain value, and one
in which
the state fraction drops systematically with time.
In both cases the simulated pulsar begins with a $\nu$ value of
1~Hz, which is constantly decreasing due to a $\dot{\nu}$ value.
The standard deviation of the uncertainty of the simulated TOAs is set
at 100~$\mu$s. Details of the simulation parameters
  are presented in Section~\ref{simulation}.
\begin{table*}
  \centering
  \caption{Variable parameters in the pulsar state-changing
    simulation.}
  \label{sim_table}
  \begin{tabularx}{0.8\textwidth}{ X l l } 
    \hline
    Description & Parameter & Value\\
    \hline
    Initial rotation frequency of the pulsar & $\nu_{0}$ & 1
    Hz\\
    Difference in the spindown rate of the two states &
    $|\Delta\dot{\nu}|$ &  10$^{-18}$ to 10$^{-14}$
    s$^{-2}$\\
    Stochastic state fraction standard deviation &
    $\sigma_{\rm SF}$ & 0.01 to 0.1\\
    Systematic rate of state fraction drop & - &
    0.005 to 0.05 per year\\
    Uncertainty of TOA measurements &
    $\sigma_{\rm TOA}$ & $1\times10^{-4}$
    seconds\\
    Time between possible state changes &
    - & 1 hour\\
    Time between $\nu$ and SF
    measurements & - & 15 days\\
    Simulation length & - & 1
    and 2 years\\
    \hline
  \end{tabularx}
\end{table*}
\subsection{Stochastic Changes in State Fraction}
\label{stochastic}
\begin{figure*}
  \centering
  \includegraphics[width=143mm]{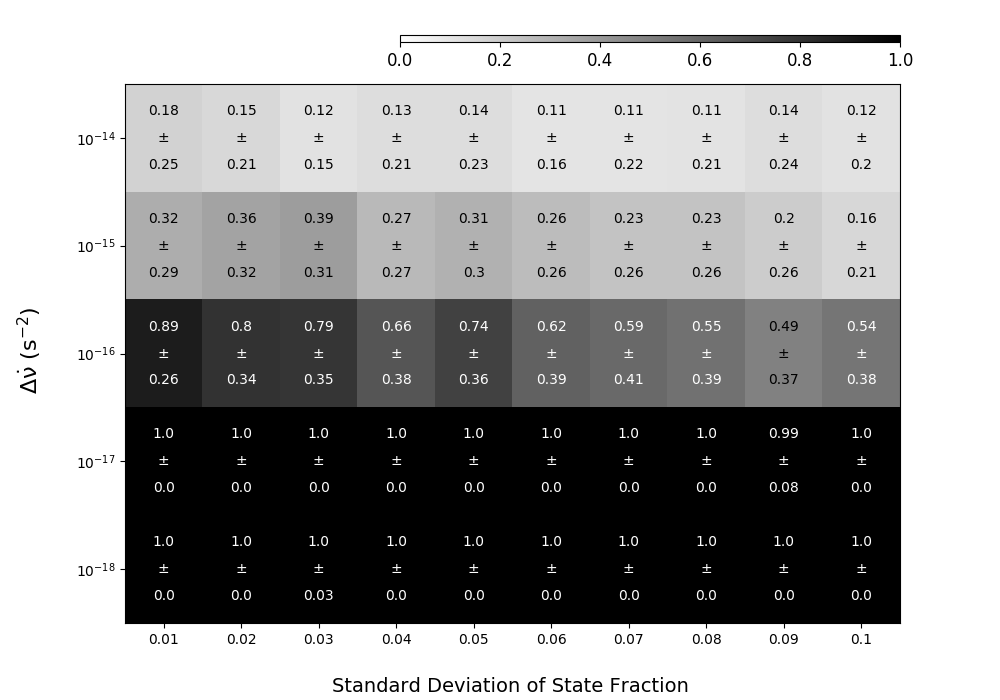}
  \caption[]{The correlation of $\dot{\nu}$ and state fraction over
    a
    range of parameters for simulations of one-year continuous
    observations. For each permutation of parameters the mean
    and
    standard deviation of 100 $p$-values is shown. A smaller
    $\bar{p}$-value reflects a lower probability that
    non-correlated data would show at least an equal
    level of correlation and produces a lighter
    grey-scale square.
    The realisations in which the GP modelled
    $\dot{\nu}$ as
    having no change over the duration of the
    simulation necessarily
    produce a Pearson correlation
    coefficient of 0 and, therefore,
    $p$-value =
    1.0. If all 100 contributing models
    have flat $\dot{\nu}$ values,
    then $\bar{p}$-value = 1.0 and
    the standard deviation will be
    0.0, as seen in
    some regions of parameter
    space with small
    $|\Delta\dot{\nu}|$
    values.}
  \label{stochastic1}
\end{figure*}

\begin{figure*}
  \centering
  \includegraphics[width=143mm]{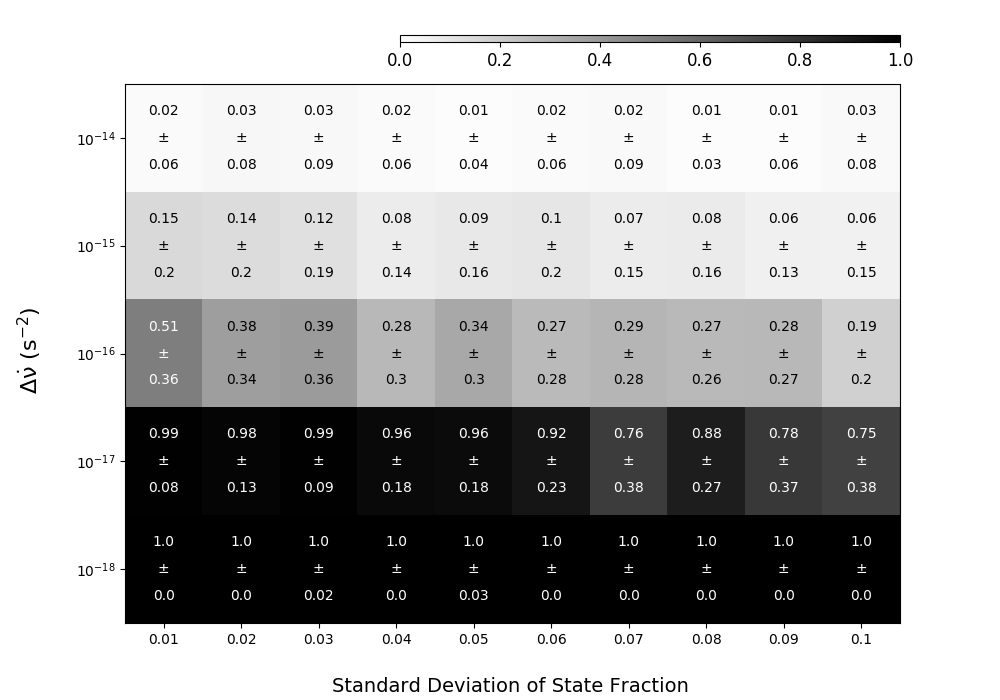}
  \caption[]{As Figure~\ref{stochastic1}, but for simulations of
    two-year continuous observations.}
  \label{stochastic2}
\end{figure*}

In one version of the simulation, we make the assumption that the
underlying state fraction of the pulsar has a mean value (which we set
to 0.5) and a standard deviation which we vary between 0.01 to 0.1
(holding the value fixed for the length of the simulated
experiment). We will see that if we draw the state fraction from a
distribution with a standard deviation too much above or below these
values, then identifying the different $\dot{\nu}$ values of a state
changing pulsar becomes either impossible or trivial respectively
(over most of our chosen range of $|\Delta\dot{\nu}|$ values). The
value of the
underlying state fraction stays fixed for 15 days; the period over
which $\dot{\nu}$ and the observed state fractions are evaluated. The
simulated pulsar has the opportunity to change between a nulling or
emitting state every hour of the simulation. This is determined by the
generation of a random number, weighted by the value of the underlying
state fraction at that point in the simulation. The frequency with
which the state of the pulsar is permitted to change has an
effect in terms of the standard deviation of the observed state
fraction in any 15-day evaluation period: $\sigma_{SF} \propto
\sqrt{T}$,
where $T$ is the nulling/emitting timescale. This is discussed
further in Section~\ref{disc}. The opportunity to change every
simulated hour was chosen to find a balance between simulating a
realistic state-changing pulsar and short computation time.
\\
As inferred in Section~\ref{simulation}, another important indicator
of whether different spindown states can be detected is how different
the $\dot{\nu}$ values of the two states are. For this simulation,
$|\Delta\dot{\nu}|$
has values between 10$^{-18}$ and 10$^{-14}$ s$^{-2}$.
These values are reasonable when considering the $|\Delta\dot{\nu}|$
values observed in known state switching and intermittent pulsars
\citep{2006Sci...312..549K,2010Sci...329..408L,2012ApJ...746...63C,2012ApJ...758..141L,2014ApJ...780L..31B,2016MNRAS.456.1374B,2017ApJ...834...72L}.\\
The pulsar is simulated to be observed for one year and also for two
years. As the observed state fraction and average $\dot{\nu}$ value
are measured every 15 days, the simulation generates 24 and 48 pairs
of data points for the one- and two-year simulations respectively. The
Pearson correlation coefficient of the pairs is then calculated along
with a $p$-value, which indicates the probability that no correlation is detected.
Figures~\ref{stochastic1} (one year) and \ref{stochastic2} (two years)
show the mean $p$-values ($\bar{p}$) for a range of state fractions
and
$|\Delta\dot{\nu}|$. For each location in parameter space in these
figures, 100 $p$-values
were calculated and the resulting $\bar{p}$-value and standard
deviation are shown.
\subsection{Systematic Changes in State Fraction}
\label{systematic}
\begin{figure*}
  \centering
  \includegraphics[width=143mm]{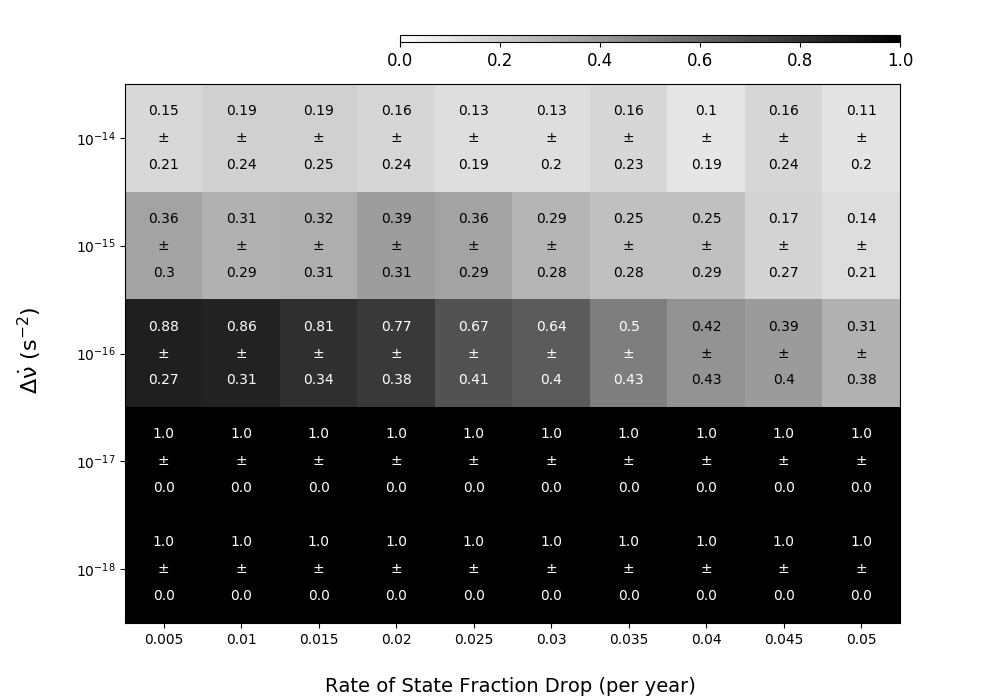}
    \caption[]{As Figure~\ref{stochastic1}, but for simulations of
      one-year continuous observations during which the state fraction
      systematically drops.}
    \label{systematic1}
\end{figure*}

\begin{figure*}
  \centering
  \includegraphics[width=143mm]{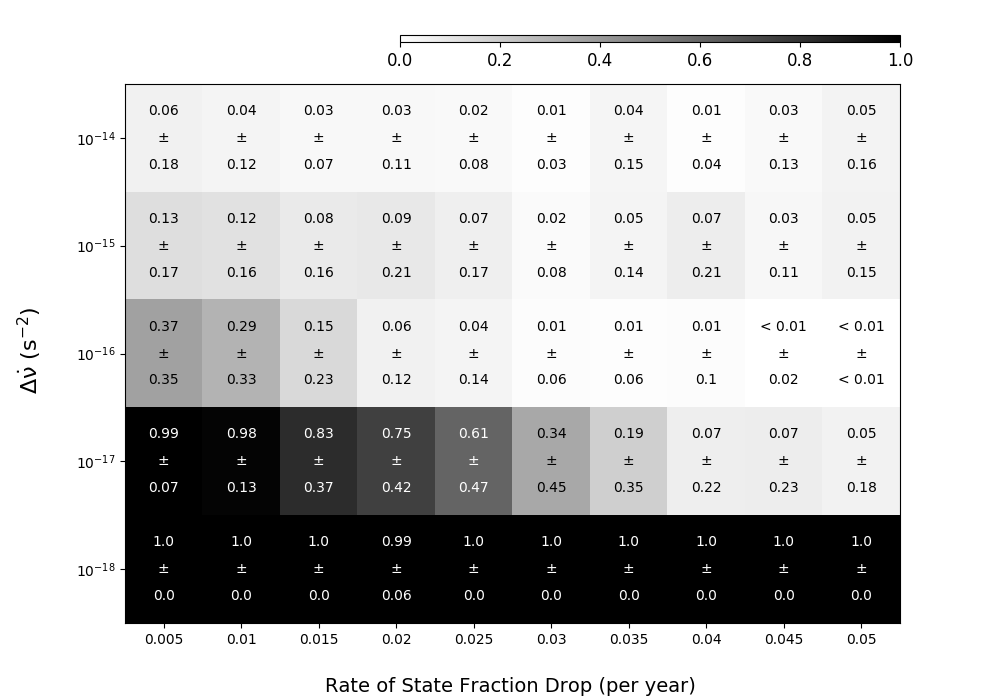}
  \caption[]{As Figure~\ref{stochastic1}, but for simulations of
    two-year continuous observations during which the state
    fraction systematically drops.}
  \label{systematic2}
\end{figure*}

In a second version of the simulation, we make the assumption that the
underlying state fraction systematically drops over time; we explore
the parameter space over which the drop rate is between 0.005 and 0.05
per
year. This range is informed by our observations of PSRs~J1701$-$3726
and J1727$-$2739; 0.005 per year amounts to a drop in state fraction
of $\sim$ 5\% over the span between our 2004 and 2014 nulling fraction
calculations. It is possible for a change of this magnitude to be
hidden in PSRs~J1701$-$3726 and J1727$-$2739 by measurement
uncertainties and statistical fluctuations. Values above our upper
value of 0.05 per year are considered unrealistically large. All other
parameters are also unchanged from the stochastic simulation including
$|\Delta\dot{\nu}|$ which is again simulated between 10$^{-18}$ and
10$^{-14}$ s$^{-2}$. The results from the one and two year simulations
are shown in Figures~\ref{systematic1} and \ref{systematic2}
respectively.
\section{Discussion}
\label{disc}
\begin{figure*}
  \centering
  \includegraphics[width=165mm]{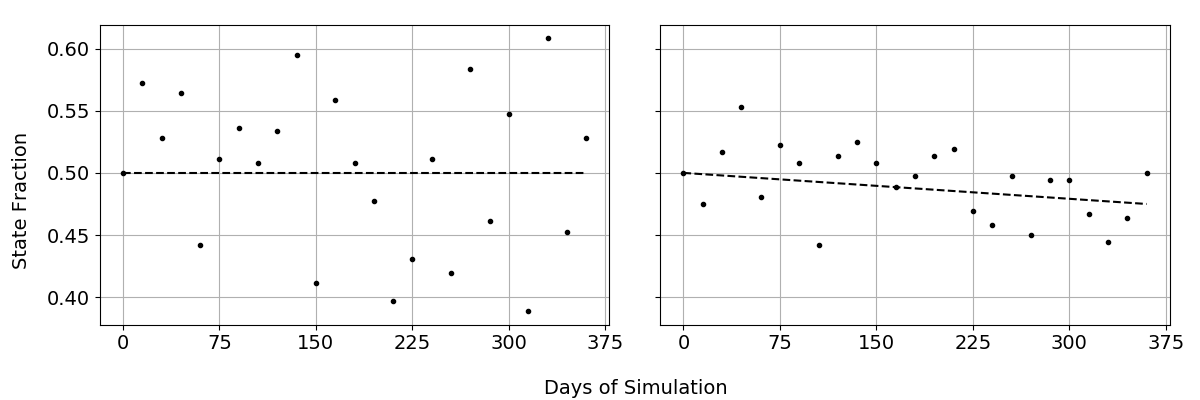}
  \caption[]{A comparison of stochastically and systematically
    varying
    state fractions. The left panel shows stochastic evolution,
    with
    $\sigma_{\rm SF} = 0.05$. The underlying mean value is shown
    by the
    dashed line. The right panel has a systematic trend; the
    underlying rate of state fraction drop is 0.025 per year, as
    shown by
    the dashed line. The state fractions in the left panel
    correlates
    with their resulting $\dot{\nu}$ values to a level of
    $p=0.1132$. For the
    right panel $p=0.1139$.}
  \label{compare}
\end{figure*}

The state-changing simulation shows us that our best opportunity to
infer different $\dot{\nu}$ states in nulling or mode-changing
pulsars,
is by observing those with high $|\Delta \dot{\nu}|$ values and with a
state fraction that is highly variable or displays significant
systematic changes.
If the intrinsic state fraction of a pulsar has a steady mean around
which it varies stochastically with time, then a large variance of the
state fraction
improves our sensitivity to the detection of distinct $\dot{\nu}$
states.
The situation is not so simple when we consider a pulsar with a
systematically varying state fraction.
There is now a trade-off
between a state fraction with a large variance and one with smaller
variance
which more faithfully follows the systematic trend. The optimum
variance of state fraction in each systematic case is dependent on the
nature of
the trend.
This is illustrated in Figure~\ref{compare} which compares the
evolution
of stochastically and systematically varying state fractions. Both
simulated data sets show a similar level of correlation between the
state fraction
and $\dot{\nu}$, with each having a final $p$-value of 0.11.

A pulsar having similar properties to those in our simulation and
a $|\Delta\dot{\nu}|$ value of at least $10^{-14}$ s$^{-2}$
will allow us to detect correlation between state fraction and
$\dot{\nu}$ with 95\% confidence within a two-year
observing campaign.
This is equivalent to a pulsar with $\dot{\nu} = -10^{-13}$ s$^{-2}$
which changes by 10\% between emission states.
Although above average, there are still many known
radio pulsars with $|\dot{\nu}|$ $\geq$ $10^{-13}$ s$^{-2}$. Of
course, in pulsars
with even higher values of  $\dot{\nu}$, a lower percentage change is
needed to satisfy
our $|\Delta\dot{\nu}|$ requirement of at least $10^{-14}$ s$^{-2}$
when the state switch occurs.
If the variability of the state fraction is very high, or
if it changes in a systematic rather than a stochastic way, then any
correlation between state fraction and $\dot{\nu}$ may be confidently
seen even
when $|\Delta\dot{\nu}|$ is lower than $10^{-14}$ s$^{-2}$.

\citet{2018MNRAS.475.5443S} show that the transitions only become
reliably detectable when they occur on timescales greater than
approximately a month. They also show that using changes in a pulsar's
emission to provide information about the transition epoch (assuming
rotation-emission correlation in the pulsar) is advantageous for
finding transition parameters when the $\dot{\nu}$ jumps are low
amplitude and closely spaced in time. Although we rely on statistical
rather than direct measurement techniques, in some sense our work is
an extrapolation of these concepts; we are able to detect rotational
changes that occur right down to the shortest timescales (the pulse
period) and our ability to do so is completely reliant on information
provided by the continuous monitoring of the emission state of the
pulsar.
\subsection{Caveats}
\label{caveats}
\begin{itemize}
\item
  When setting 100~$\mu$s as a typical level of TOA measurement
  uncertainty, we only considered template-fitting errors due to
  radiometer noise. However, phenomena such as \emph{pulse jitter}
  \citep{1985ApJS...59..343C} are known to be present in some
  pulsars. This is the stochastic, broadband, single-pulse variations
  that are intrinsic to the pulsar emission process and affect the
  shape
  of the integrated pulse profile. The presence of jitter would
  increase
  the TOA measurement uncertainty and hinder the detection of a
  correlation between $\dot{\nu}$ and state fraction. However, TOA
  uncertainty due to jitter
  \begin{equation}
    \sigma_{J} \propto \frac{1}{\sqrt{n}},
  \end{equation}
  where n is the number of pulses that make up the pulse profile used
  to
  calculate the TOA.
  As we are proposing to calculate TOAs over a 15~day span, the many
  integrated pulses ensure that this uncertainty is small. From
  Equation~5 of \citet{2010arXiv1010.3785C} we calculate that the TOA
  measurement error due to jitter for a typical pulsar with $\nu$ =
  1~Hz, calculated over 15~days is $\sim$ 2~$\mu$s. When this jitter
  uncertainty is added in quadrature to the template-fitting
  uncertainties, the latter will dominate. Timing uncertainty induced
  by
  jitter can, therefore, largely be ignored in our proposed
  experiment.

\item
  The state-changing simulation does not include the injection of the
  timing irregularity known as $\emph{timing noise}$. This is a term
  given to the unexplained, quasi-periodic wander from the modelled
  rotational behaviour of a pulsar. There have been numerous processes
  proposed to explain timing noise, such as the presence of an
  asteroid
  belt \citep{2013ApJ...766....5S}, or planetary systems
  \citep{1999ApJ...523..763T}. Both \citet{2006Sci...312..549K} and
  \citet{2010Sci...329..408L} showed that timing noise can be produced
  by unmodelled magnetospheric state changes that simultaneous affect
  a
  pulsar's emission and rotation (in intermittent pulsars and
  state-switching pulsars respectively). If the short-term emission
  variability in nulling and mode-changing pulsars is also accompanied
  by spindown rate changes, then timing noise will also be intrinsic
  to
  these pulsars and hence may naturally emerge from our simulations.

\item
  The $|\Delta\dot{\nu}|$ values in the simulations were based on
  state-switching pulsars \citep{2010Sci...329..408L} which have
  reported fractional changes in
  $\dot{\nu}$ of between approximately 1-10\%, and intermittent
  pulsars
  \citep{2006Sci...312..549K,2012ApJ...746...63C,2012ApJ...758..141L,2017ApJ...834...72L}
  which have fractional changes up to around 150\%. At present, we do
  not know if the
  fractional $\dot{\nu}$ changes in nulling and mode-changing
  pulsar will be comparable, if indeed they change at all.
  Although they have similar
  timescales, the radio emission in intermittent pulsars
  appears to cease completely (unlike state-switching pulsars).
  By analogy, we might expect that nulling
  pulsars may also have larger fractional $\dot{\nu}$ changes than
  mode-changing
  pulsars when their shorter timescale state changes occur.

\item
  When modelling how the pulsar state fraction changes with time, the
  variable input parameter for the simulation is (i) how the
  underlying
  state fraction changes with time. When we subsequently measure the
  output state fraction, however, the result will be a
  combination of (i) and (ii) the standard deviation of the
  measurements due to the statistics of
  finite observation length. If the underlying state fraction has an
  unchanging value, the measurements will vary around this mean; the
  standard deviation of the state fraction in this case would depend
  on
  how many state changes take place during an observation, and can be
  approximated as a binomial process. As an example, if the underlying
  state fraction of a pulsar is unchanging at 0.5, then a 15-day
  observation in which there is an hourly opportunity to switch states
  (as in our simulation) would constitute 360 trials. Therefore,
  $\sigma_{\rm STATE} = \sqrt{360 \cdot 0.5 \cdot 0.5}/360 =
  2.6\%$. If the pulsar
  was able to switch states each minute, then $\sigma_{\rm STATE}$
  drops to
  0.3\%.
  As we want to maximise the variance of state fraction in order for
  us
  to detect any correlation between state fraction and $\dot{\nu}$, it
  would be preferential for us to observe a pulsar in which the state
  changes occur on as long a timescale as possible.
  Conversely, when considering a pulsar in which the state changes
  occur
  on timescales less than an hour, our results matrices
  (Figures~\ref{stochastic1} to \ref{systematic2}) will be optimistic,
  especially in regions where the standard deviation of the underlying
  state fraction (Figures~\ref{stochastic1} and \ref{stochastic2}) or
  rate of state fraction drop (Figures~\ref{systematic1} and
  \ref{systematic2}) is low.
\end{itemize}
Even in the most pessimistic case, in which a continuous monitoring
campaign of a state-changing pulsar does not yield any correlation
between $\dot{\nu}$ and state fraction, this would allow an upper
limit to be placed on $|\Delta\dot{\nu}|$. In addition to this, such a
campaign will produce a unique data set and provide information
regarding  how the state fraction of nulling or mode-changing pulsars
evolves over timescales from days to years. To carry out this experiment in practice, the observing instrument need only monitor continuously
for as long as it takes to make a precise measurement of $\dot{\nu}$ (around 15 days) and the corresponding state fraction for the observation span. Each such pair of data points can be recorded in this way, with no requirement for the observing spans to be contiguous. Therefore, the observing instrument does not need to be employed continuously for many months. In principle, a bright circumpolar
pulsar could be observed with a relatively high $S/N$ using a sub-array of a radio interferometer rather than a dedicated single dish telescope. Any instrument must be sufficiently sensitive to obtain the necessary $S/N$ to obtain precise TOAs and also to be able to distinguish between different emission modes.
\section{Conclusions}
\label{conc}
We have simulated the parameters over which a continuous monitoring
campaign of a state-changing radio pulsar could reveal distinct
spindown rates in each emission state. All other things being equal,
the simulation results have shown us that the crucial parameters for
success are (i) a long monitoring campaign, (ii) a state fraction that
is either highly variable or follows a significant systematic trend
and (iii) a large difference between state spindown rates. The latter
will not be known before the experiment takes place, and (ii) may only
be poorly constrained at best; if a pulsar is known to have a
predictable systematic state fraction, then it may be possible to
forego continuous monitoring altogether and just take enough
observations to compare $\dot{\nu}$ against the predicted state
fraction changes. Assuming no knowledge of (ii), in order to maximise
our chances of success in revealing distinct rotational states, we
would ideally monitor a bright, circumpolar nulling pulsar with a high
rate of spindown, a long timescale for nulls and a nulling fraction
close to 50\% to maximise the statistical variance of the nulling
fraction.
\section*{Acknowledgments}
P.R.B. is supported by Track I award OIA-1458952 and is a member of
the NANOGrav Physics Frontiers Center, which is supported by NSF award
number 1430284. The Parkes radio telescope is part of the Australia
Telescope National Facility which is funded by the Commonwealth of
Australia for operation as a National Facility managed by CSIRO. Data
taken at Parkes is available via a public archive. We thank the referee
Patrick Weltevrede for valuable comments that helped to improve the text.\\

\bibliographystyle{mnras}
\bibliography{bkj_2019.bib}

\bsp
\label{lastpage}
\end{document}